\documentclass[journal=apchd5,manuscript=letter]{achemso}
\usepackage{epstopdf}
\usepackage{amsmath}
\usepackage{bm}
\usepackage{braket}

\author{Jordi Llusar}
\affiliation{Departament de Qu\'{\i}mica F\'{\i}sica i Anal\'{\i}tica,
Universitat Jaume I, E-12080, Castell\'o de la Plana, Spain}

\author{Juan I. Climente}
\affiliation{Departament de Qu\'{\i}mica F\'{\i}sica i Anal\'{\i}tica,
Universitat Jaume I, E-12080, Castell\'o de la Plana, Spain}
\email{climente@uji.es}

\title{Nature and Control of Shakeup Processes in Colloidal Nanoplatelets}


\begin{document}

\begin{tocentry}
\includegraphics[height=3.5cm]{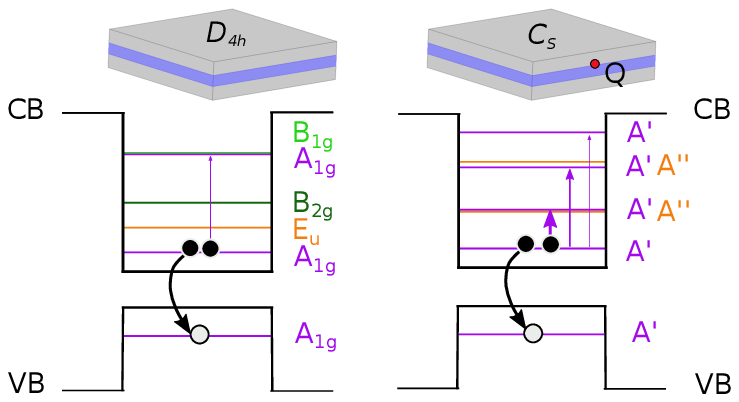}
\end{tocentry}

\begin{abstract}
Recent experiments suggest that the photoluminescence line width of CdSe and CdSe/CdS 
nanoplatelets (NPLs) may be broadened by the presence of shakeup (SU) lines from 
negatively charged trions. 
We carry out a theoretical analysis, based on effective mass and configuration
interaction (CI) simulations, to identify the physical conditions that enable
such processes. 
We confirm that trions in colloidal NPLs are susceptible of presenting 
SU lines up to one order of magnitude stronger than in epitaxial quantum wells, 
stimulated by dielectric confinement. 
For these processes to take place trions must be weakly bound to off-centered impurities,
which relax symmetry selection rules. 
Charges on the lateral sidewalls are particularly efficient to this end.
We propose that the broad line width reported for core/shell CdSe/CdS NPLs may relate
not only to SU processes but also to a metastable spin triplet trion state.
Understanding the origin of SU processes opens paths to rational design of NPLs with narrower line width.
\end{abstract}


%\section{Introduction}

Colloidal metal chalcogenide NPLs offer well defined advantages over their quantum dot 
and rod counterparts as semiconductor building blocks for optical applications.\cite{LhuillierACR,MinNR,DirollJMCc,SharmaIEEE}
Some of the most distinctive features are order-of-magnitude shorter radiative lifetimes,
which result from the strong exciton binding energies in quasi-2D systems
(Giant Oscillator Strength effect),\cite{FeldmannPRL,PlanellesACSph}
and precisely controlled thickness of the nanostructure,\cite{IthurriaNM,RiedingerNM,ChristodoulouNL,BhandariCM}
which suppresses the emission broadening due to size dispersion usually observed in dots.
These properties give rise to bright and narrow emission lines, which
is of interest for displays, lighting and lasers.\cite{DirollJMCc,SharmaIEEE}

Unfortunately, ligand passivation of surface dangling bonds is usually incomplete because of
 labile binding and steric hindrance between ligands. This can translate into significant 
non-radiative losses.\cite{TessierACS} To overcome this problem, core-only NPLs are sometimes
replaced by sandwich-like core/shell heterostructures, where top and bottom facets of the 
core material are coated with a higher band gap inorganic material. 
Typical core/shell combinations are CdSe/CdS,\cite{TessierNL,AchtsteinACS} 
CdSe/ZnS\cite{PolovitsynCM,SaidzhonovJL} and their alloys.\cite{RossinelliCM,KelestemurACS}
These heterostructures succeed in isolating the photogenerated carriers, 
which remain in and around the core, from top and bottom surfaces,
thus translating into enhanced fluorescence quantum efficiency and 
photostability.\cite{LhuillierACR,SharmaIEEE,YadavJPCc}
The shell growth has however a negative side effect, namely the systematic broadening of
the emission line width, e.g. from $\sim 35-40$ meV in CdSe NPLs to $\sim 60-80$ meV in CdSe/CdS NPLs.\cite{TessierNL,RabouwNL}
 Linewidth broadening in core/shell NPLs was initially ascribed to the presence of traps
induced upon shell coating.\cite{TessierNL} Graded interface composition was then shown to 
narrow the line width down to $\sim 55$ meV,\cite{RossinelliCM,KelestemurACS} 
but this figure is still larger than in core-only NPLs, which suggests that interface 
defects are not the only source of broadening.

To shed light into this problem, Antolinez and co-workers recently investigated the origin 
of the fluorescence line width broadening in CdSe/CdS NPLs by means of single-particle spectroscopy.\cite{AntolinezNL} 
They observed that individual NPLs present a series of 2 to 4 narrow peaks split from each other by $\sim 10$ meV.
Altogether, the peaks fit well the asymmetric lineshape of ensemble NPLs at cryogenic temperatures.\cite{TessierNL}
A similar feature was soon after reported in core-only CdSe NPLs.\cite{AntolinezNL2}
The nature of these peaks was tentatively ascribed to SU processes of negative trions (X$^-$).
These are partly radiative Auger processes, whereby an electron-hole pair recombines radiatively but 
transfers part of its energy to the remaining electron by exciting it into a higher single-electron level
(in-plane excitation).
%Shake-up processes are enabled in weakly confined structures, where the initial state 
%(trion) wave function is defined by Coulomb interactions but that of the final state 
%(single-electron) is not.
 They have been previously reported in epitaxial quantum wells\cite{NashPRL,FinkelsteinPRB,BryjaPRB,DzyubenkoPRB} 
and self-assembled quantum dots\cite{PaskovPE} under the magnetic fields, corresponding 
to inter-Landau level excitations of the excess carrier.\cite{HawrylakPRB}
 Clarifying the role of SU processes in the emission of colloidal NPLs is then a desirable step
to fully understand and control the emission line width, which would be advantageous for optical 
applications.

In this work, we analyze the possible occurence of SU processes in colloidal CdSe-based NPLs
from a theoretical perspective. The goal is to determine which physical conditions enable 
these processes. % and how to control them.
To this end we use effective mass models and full CI simulations,
which provide an intuitive description of the underlying physics.
 We shall confirm that at least one intense SU replica can be expected for $X^-$
upon electron-hole recombination, in both core-only and core/shell NPLs, 
corresponding to the excitation of the remaining electron into a higher orbital 
with the same symmetry as the ground state.
For this to take place, the trion must be weakly bound to an off-centered acceptor impurity.
The role of the impurity is to lower the system symmetry, thus relaxing selection rules, 
and to stimulate electron-electron repulsion (quench electron-hole attraction) in the 
ground orbital. %Surface traps on the lateral sidewall are found to be particularly efficient at this regard.
 By doing so, SU peaks can reach intensities exceeding 10\% of the fundamental (band edge, fully radiative) transition. 
This is one order of magnitude higher than in epitaxial quantum wells, which can be rationalized
from the stronger Coulomb interactions, which result from the pronounced dielectric confinement,
and the presence of lateral sidewalls, which are prone to surface traps.
 We discuss connections with experiments in the literature and propose potential 
 strategies to suppress these processes.

\section{Results}

We analyze the emission spectra of trions in core-only and core/shell NPLs. 
Negative trions are studied unless otherwise noted, as it is the
most frequently reported species in these structures, but the conclusions do not
depend on the sign of the charged exciton (see Fig.~S2 in the supporting information, SI).
Once the general behavior of SU processes in these systems is understood, 
we discuss how our conclusions fit the interpretation of different experimental 
observations and the practical implications of our findings.

\subsection{Core-only NPLs}

\begin{figure}
	\centering
		\includegraphics[width=8cm]{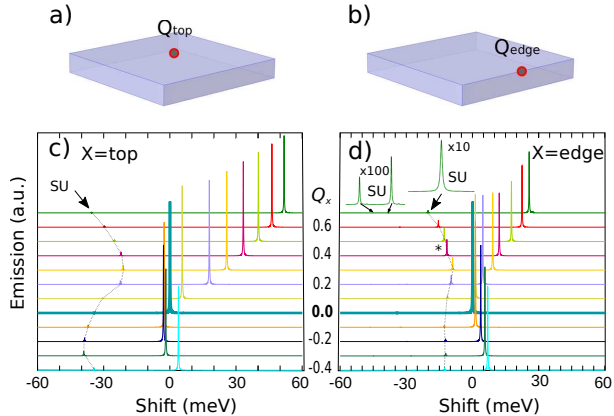}
\caption{
(a,b) Schematics of core-only NPLs with different location of the surface charge. 
(c,d) Corresponding X$^-$ emission spectrum for charge strength $Q=Q_X\,e$. 
The arrows point at the SU satellites 
(dotted lines are guides to the eyes). The highest SU peak is observed 
for off-centered acceptor charges weakly bound to the trion ($Q_{edge} = 0.4$, 
marked with a star in (d)). 
The spectra are normalized to the intensity of the fundamental transition
at $Q_X=0$, and offset vertically for clarity.
The insets for $Q_{edge}=0.7$ in (d) show amplified SU peaks.
}\label{fig1}
\end{figure}

We start by studying core-only CdSe NPLs. The NPLs are chosen to have $4.5$ monolayer (ML) thickness 
and a lateral size of $20 \times 20$ nm$^2$, for similarity with the core dimensions 
of Ref.~\cite{AntolinezNL}. They have a pronounced dielectric mismatch with the organic 
environment, which we model with $\epsilon_{in}=6$ and $\epsilon_{out}=2$ as dielectric 
constants inside and outside the NPL, unless otherwise stated.\cite{Sadao_book,AchtsteinNL}
 The presence of few-meV spectral jumps in photoluminescence experiments\cite{AntolinezNL} 
suggests that the trion is subject to the influence of carriers temporarily trapped on 
the surface.\cite{RabouwNL,BeylerPRL} 
To model this phenomenon, a fractional point charge is placed on the surface, 
with charge $Q=e \, Q_X$ ($|Q_X| \leq 1$ and $e$ the full electron charge).
The fractional value of $Q_X$ accounts for the screening of trapped charged (e.g. hole)
by the trap defect itself (e.g. surface dangling bond).\cite{CalifanoNL}
Two scenarios are considered: a charge centered on the top facet ($Q_{top}$) 
and an off-centered charge, located along the edge of a lateral facet ($Q_{edge}$). 
The latter setup is suggested by studies showing that edge and vertex atoms in CdSe 
structures have weaker binding to oleate ligands.\cite{DrijversCM}
The two systems are represented in Figure \ref{fig1}a and \ref{fig1}b.
The corresponding emission spectra are shown in Fig.~\ref{fig1}c and \ref{fig1}d. 
The figure reveals a number of important observations.
(i) In the absence of surface charge ($Q_X=0$, thick lines),
only the fundamental transition shows up, with no sizable SU replica. 
(ii) Charges on the top facet induce SU peaks (see arrow in Fig.~\ref{fig1}{c}),
but their strength is two orders of magnitude smaller than that of the 
fundamental transition (main line). This is similar to the case of epitaxial quantum wells.\cite{NashPRL,FinkelsteinPRB,BryjaPRB,DzyubenkoPRB}
(iii) Stronger SU replica are however obtained for charges located on the 
lateral sidewall, provided the charge is attractive (acceptor impurity) 
and binding to the trion is moderately weak, see Fig.~\ref{fig1}d.
For $Q_{edge}=0.4$ (marked with a star in the figure),
the SU peak reaches $\sim 25$\% of the main peak height.
This ratio is about 20 times higher than in epitaxial quantum wells,
and it holds despite the Giant Oscillator Strength enhancing the band edge 
recombination,\cite{FeldmannPRL,PlanellesACSph,IthurriaNM,AchtsteinNL}
which suggests that SU satellites also benefit from this phenomenon.
For $Q_{edge}>0.4$, however, the SU peak intensity is lowered again
and the energy splitting (redshift) with respect to the main line increases.
Second and third SU lines are built for strong surface charges 
(see inset in Fig.~\ref{fig1}d at $Q_{edge}=0.7$), 
but their magnitude is negligible.
We have also explored different locations of the charge, obtaining intermediate
results between those shown in Fig.~\ref{fig1} (see Fig.~S3 in SI). 
These results point out the potentially significant role of lateral sidewalls, 
which are characteristic feature of colloidal quantum wells as compared to epitaxial ones, 
in obtaining high SU peaks.

\begin{figure}
	\centering
		\includegraphics[width=13cm]{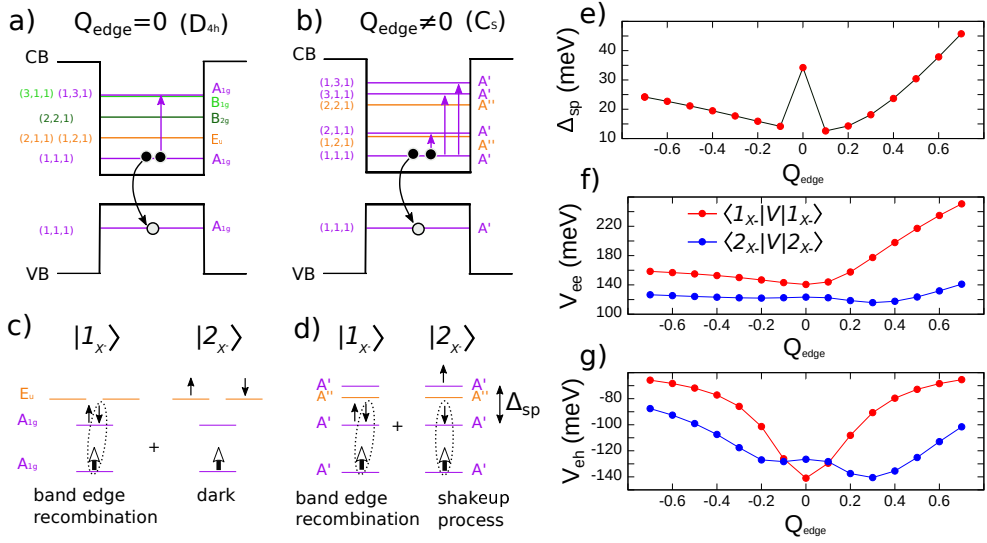}
\caption{
(a,b) Sketch of SU processes in NPLs with (a) and without (b) an edge charge.
Labels on the left are $(n_x, n_y, n_z)$ quantum numbers
for the (independent particle) energy levels.
Labels on the right are the corresponding irreducible representation. 
The surface charge lowers the point group symmetry, 
from $D_{4h}$ to $C_s$, lifting degeneracies and enabling
new channels for SU transitions (vertical arrows).
(c,d) Two main configurations $|m_{X^-}\rangle$ in the CI expansion of $|GS_{X^-}\rangle$,
with and without edge charge. Thin (thick) arrowsheads denote electron (hole) spin.
Only when $Q_{edge}\neq 0$ a SU process is expected.
(e) Energy splitting between $|1_{X^-}\rangle$ and $|2_{X^-}\rangle$ at an 
independent particle level.
(f) average value of electron-electron repulsion and (g) electron-hole
attraction in configurations $|1_{X^-}\rangle$ and $|2_{X^-}\rangle$.}\label{fig2}
\end{figure}

To gain understanding on the origin of strong SU peaks when trions bind
to lateral surface acceptors, beyond the full numerical calculation
of Fig.~\ref{fig1}, in Fig.~\ref{fig2}a and \ref{fig2}b we 
compare sketches of the SU processes,
in the absence and presence of an attractive edge charge.
Within effective mass theory, the conduction band 
and valence band energy levels of (non-interacting) 
electrons and holes can be described as particle-in-the-box states, 
with quantum numbers $(n_x,n_y,n_z)$. 
It is useful however to label the states by their symmetry 
(irreducible representation).
When $Q_{edge}=0$, because the NPL has squared shape, the 
point group is $D_{4h}$. When $Q_{edge}\neq 0$, the electrostatic
potential yields a symmetry descent to $C_s$.
As a consequence, degeneracies are lifted and additional
states with the same symmetry as the ground orbital ($A'$)
are obtained. This is important because after electron-hole
recombination, the excess electron can only be excited to
an orbital with the same symmetry as the initial one
(vertical arrows in Fig.~\ref{fig2}a and \ref{fig2}b).
Therefore, lowering the system symmetry opens new channels
for SU processes. Furthermore, these can involve low-energy 
orbitals, which have fewer nodes and will then have larger
overlap with the trion ground state, as we shall see below. 
Both the number and the intensity of the SU processes 
are in principle enhanced. 
By contrast, a centered charge on the top surface barely affects 
the system symmetry, which remains high ($C_{4v}$), and
SU processes are only slightly stronger than in the $Q_{edge}=0$ case.

The qualitative reasoning above can be substantiated with a 
CI formalism on the basis of independent particle (non-interacting) 
electron and hole states, which has the additional advantage of giving 
intuitive insight on how Coulomb interactions affect the likelihood of SU processes.
We consider that the transition rate from the trion ground state 
$|GS_{X^-}\rangle$ to an electron spin-orbital $|f_e\rangle$, 
is proportional to:\cite{Pawel_book}
\begin{equation}
P_{GS \rightarrow f} = \left| \langle f_e | \,{\hat P}\, | GS_{X^-} \rangle \right|^2 .
\label{eq:trans}
\end{equation}
\noindent ${\hat P}$ is the dipolar transition operator,
${\hat P} = \sum_{i_e,i_h} \, \langle i_e | i_h \rangle \, e_{i_e}\,h_{i_h}$,
where $e_{i_e}$ and $h_{i_h}$ are annihilation operators for independent electron
and hole spin-orbitals $|i_e\rangle$ and $|i_h\rangle$, respectively.
We describe the trion ground state with a CI expansion,
\begin{equation}
|GS_{X^-} \rangle  = \sum_m c_m \, | m_{X^-} \rangle, 
\end{equation}
\noindent where $|m_{X^-}\rangle$ is a trion configuration:
%
%\begin{equation}
$|m_{X^-} \rangle  = e_{r_e}^\dagger e_{s_e}^\dagger |0 \rangle_e \, h_{t_h}^\dagger |0 \rangle_h$,
%\end{equation}
%
%\noindent 
 with $e_{r_e}^\dagger$ and $h_{t_h}^\dagger$ creator operators,
 $|0\rangle_e$ and $|0\rangle_h$ the vacuum occupation vectors of electron and hole, 
and $c_m$ the coefficient in the expansion.
% which generate  single-electron and single-hole spin orbitals, $e_r^\dagger |0\rangle_e=|\phi_r\rangle_e$ and $h_{r'}^\dagger |0\rangle_h=|\phi_{r'}\rangle_h$.
%on the Hilbert space of electron spin-orbitals, $|0\rangle_e$, 
%and hole spin-orbitals $|0 \rangle_h$, respectively.
 Inserting ${\hat P}$ and $|GS_{X^-}\rangle$ into Equation (\ref{eq:trans}), 
 one obtains:
\begin{equation}
P_{GS \rightarrow f} = 
\left| \sum_m  c_m \, \left( \langle r_e | t_h \rangle\, \delta_{f_e\, s_e} 
- \langle s_e | t_h \rangle \delta_{f_e\, r_e} \right) \right|^2.
\label{eq:trans2}
\end{equation}
\noindent In SU processes, $|f_e\rangle$ is an excited spin-orbital. 
It then follows from Equation (\ref{eq:trans2}) that such a transition will 
only take place if $|GS_{X^-}\rangle$ contains at least one configuration $|m_{X^-}\rangle$
in the CI expansion where one electron is in the excited spin-orbital 
and the other electron has finite overlap with the hole ground state 
($|s_e\rangle = |f_e\rangle$ and $\langle r_e | t_h \rangle \neq 0$
or $|r_e\rangle = |f_e\rangle$ and $\langle s_e | t_h \rangle \neq 0$). 
The larger the weight of this configuration, $|c_m|^2$, the more likely the SU process.
It is worth noting that in the strong confinement limit, the trion ground state is well
described by a single configuration where all carriers are in the lowest-energy spin-orbitals
(configuration $|1_{X^-}\rangle$ in Fig.~\ref{fig2}c and \ref{fig2}d).
That is, $c_1 \approx 1$ and $c_m \approx 0$ for $m > 1$. 
SU transitions are then forbidden, which is why SU peaks are rarely reported 
in quantum dots.
On the contrary, in systems where Coulomb interaction energies exceed quantum
confinement energies, the CI expansion contains mono- and biexcitations of electrons.
SU processes are then enabled.
Colloidal NPLs constitute an ideal system at this regard, because they combine weak
confinement in the lateral direction with strong Coulomb 
interactions.\cite{RajadellPRB,RichterPRM}
Hereafter, we refer to this condition ($c_m \neq 0$ for $m > 1$) as Coulomb admixture.

The role of Coulomb correlation and symmetry breaking in activating SU processes
can be illustrated, in the simplest approximation, by considering the two lowest-energy 
configurations of the
trion ground state, 
\begin{equation}
|GS_{X^-}\rangle \approx c_1 |1_{X^-}\rangle + c_2 |2_{X^-}\rangle. 
\end{equation}
In Fig.~\ref{fig2}c and \ref{fig2}d we depict such configurations
in the absence and presence of an edge charge, respectively.
These can be expected to be the two most important configurations in the full CI expansion.
Notice that the two configurations must have the same symmetry, 
for Coulomb interaction to couple them. %Since the lowest energy configuration,
%$|1_{X^-}\rangle$, is totally symmetric, so must be $|2_{X^-}\rangle$.
Because the lowest-energy configuration, $|1_{X^-}\rangle$, is always totally symmetric,
so must be $|2_{X^-}\rangle$.
Thus, when $Q_{edge}=0$ ($D_{4h}$ group), the electronic configuration of $|1_{X^-}\rangle$ is
$[A_{1g}^2]_e \, [A_{1g}]_h$, and that of $|2_{X^-}\rangle$ is $[E_{u}^2]_e \, [A_{1g}]_h$. 
%Then, when $Q_{edge}=0$, $|2_{X^-}\rangle$ involves a biexcitation, 
%with the two electrons in $E_u$ orbitals, see Fig.~\ref{fig2}c. 
%Their recombination with the hole, which stays in a $A_{1g}$ orbital, is then 
The recombination of the $E_u$ electrons with the hole, which stays in a $A_{1g}$ orbital, is then 
symmetry forbidden ($\langle r_e | t_h \rangle = \langle s_e | t_h \rangle = 0$ in Eq.~(\ref{eq:trans2})).
By contrast, when $Q_{edge} \neq 0$ ($C_s$ group), $|2_{X^-}\rangle$ is formed by a monoexcitation where
one electron is placed in the $(n_x,n_y,n_z)=(2,1,1)$ orbital, which also has $A'$ symmetry, resulting
in an electronic configuration $[A'\,A']_e \, [A']_h$ (see Fig.\ref{fig2}d). %, where one electron is placed in a low-energy excited orbital with $A'$ symmetry,
 %see Fig.~\ref{fig2}d.
 The hole can then recombine with the ground orbital electron, as both have $A'$ symmetry 
 ($\langle r_e | t_h \rangle \neq 0$ or $\langle s_e | t_h \rangle \neq 0$ in Eq.~(\ref{eq:trans2}))
and leave the excited electron as the final state. 
This constitutes a SU process. 
%The larger $|c_2|^2$, the stronger the resulting peak.
 Because both SU and fundamental transition rely on the recombination of the same electron-hole pair
 (same overlap integral, e.g. $\langle r_e | t_h \rangle$), the ratio between SU and fundamental
 radiative rates can be approximated as:
 \begin{equation}
	 \frac{ P_{GS \rightarrow (2,1,1)_e} }{ P_{GS \rightarrow (1,1,1)_e} } \approx 
	 \frac{|c_2|^2}{|c_1|^2}.
 \end{equation}
 \noindent i.e. it is set exclusively by the degree of Coulomb admixture.

One can guess the requirements that maximize $|c_2|^2$ by looking which
conditions favor energetically $|2_{X^-}\rangle$ over $|1_{X^-}\rangle$. 
These include: (i) small energy splitting between the two configurations, 
at an independent particle level, $\Delta_{sp}$ in Fig.~\ref{fig2}d, 
(ii) weaker electron-electron repulsion ($V_{ee}$) and 
(iii) stronger electron-hole attraction ($V_{eh}$) in $|2_{X^-}\rangle$ as compared to $|1_{X^-}\rangle$.
Figures ~\ref{fig2}e-g show that these conditions are met for moderately attractive (positive) 
charges ($Q_{edge} \sim 0.3-0.4$).
When the off-centered charge is switched on, $\Delta_{sp}$ rapidly decreases (see Fig.~\ref{fig2}e) because 
the symmetry descent turns one of the $E_u$ ($p$-like) electron orbitals into a $A'$ ($s$-like) one.
However, the surface charge brings about electrostatic confinement and hence $\Delta_{sp}$ increases again soon after.
As for inter-electron repulsion, $\langle 1_{X^-} | V_{ee} | 1_{X^-} \rangle$ 
increases more rapidly than $\langle 2_{X^-} | V_{ee} | 2_{X^-} \rangle$
(see Fig.~\ref{fig2}f) because the former involves placing the two electrons in 
identical orbitals, while the latter does not. % Fig.~\ref{fig2}f. %, while the latter involves one
 %$s$-like but one $p$-like orbital -the excited $A'$ orbital has one node-.
Last, $\langle 1_{X^-} | V_{eh} | 1_{X^-} \rangle$ is rapidly quenched (see Fig.~\ref{fig2}g) 
because it involves the ground orbitals of electron and hole --$(1,1,1)_e$ and $(1,1,1)_h$--, 
which dissociate rapidly under an external charge.
$\langle 2_{X^-} | V_{eh} | 2_{X^-} \rangle$ stays strong up to $Q_{edge} \sim 0.3$ because it involves the $(2,1,1)_e$ orbital,
which is spatially more extended and then keeps significant overlap with the $(1,1,1)_h$ hole. 
Figs.~\ref{fig2}e-f further evidence that $Q_{edge} > 0.3-0.4$ is inconvenient for SU processes, 
because the electrostatic potential increases lateral 
quantum confinement ($\Delta_{sp}$ increases) and because electrons and hole in configuration $|2_{X^-}\rangle$ are
eventually dissociated as well ($\langle 2_{X^-} | V_{eh} | 2_{X^-} \rangle$ is quenched in Fig.~\ref{fig2}g).

\begin{figure}
	\centering
		\includegraphics[width=8cm]{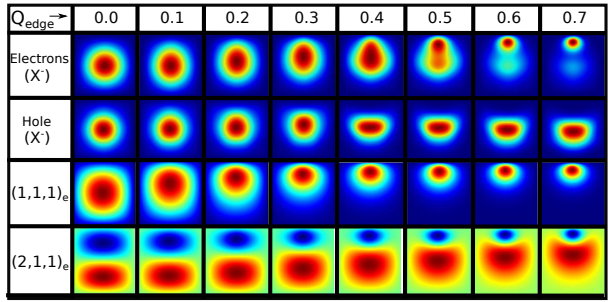}
\caption{In-plane charge density of the two electrons and hole in the X$^-$ ground state (top rows), 
and wave functions of the two lowest electron orbitals with $A'$ symmetry (bottom rows), 
as a function of the edge charge magnitude. The edge charge is located on the top edge, in this view. 
The strongest SU peak corresponds to $Q_{edge}\approx 0.4$,
when the X$^-$ electron charge density reveals a clear contribution from $(2,1,1)_e$, and the hole
is not yet fully dissociated from electrons. % the hole density still overlaps with $(1,1,1)_e$.  overlap with $(1,1,1)
	}\label{fig3}
\end{figure}

Much of the above observations can be visualized by analyzing the evolution
of charge densities and wave functions under $Q_{edge}$. 
In Figure \ref{fig3} we show the two-electron (first row)
and one-hole (second row) charge densities of $|GS_{X^-}\rangle$, 
obtained from the CI calculations of Fig.~\ref{fig1}.
The wave functions of the two lowest electron orbitals which can constitute 
configuration $|2_{X^-}\rangle$, --$(n_x,n_y,n_z)_e=(1,1,1)_e$ and $(2,1,1)_e$-- 
are also plotted (bottom rows).
At $Q_{edge}\approx 0$, the two orbitals are quasi-orthogonal. 
As a result, Coulomb interaction cannot couple configurations 
$|1_{X^-}\rangle$ and $|2_{X^-}\rangle$. %as 
%$\langle 1_{X^-} | V_{ee} | 2_{X^-} \rangle = 
%\langle 1_{X^-} | V_{eh} | 2_{X^-} \rangle = 0$.
 and $c_2 \approx 0$. This is why the two-electron charge density 
closely resembles the $(1,1,1)_e$ orbital. SU processes are not expected in this case.

At $Q_{edge} \approx 0.4$, symmetry lowering and energetic considerations 
enable efficient Coulomb coupling. 
The oval shape of the two-electron charge density reflects a significant contribution 
from $(2,1,1)_e$ to $|GS_{X^-}\rangle$ (i.e. $|c_2| > 0$). 
At the same time, the electron $(1,1,1)_e$ orbital and the hole ground state have 
sizable overlap.  This is an optimal situation for the appearance for
the transition $P_{GS \rightarrow (2,1,1)_e}$ to show up as a SU process,
according to Equation (\ref{eq:trans2}).
Further increasing $Q_{edge}$ separates the $(2,1,1)_e$ electron orbital from
the hole. Coulomb attraction is then weaker, making $c_2$ and consequently 
$P_{GS \rightarrow (2,1,1)_e}$ small again.
\begin{figure}
	\centering
		\includegraphics[width=6cm]{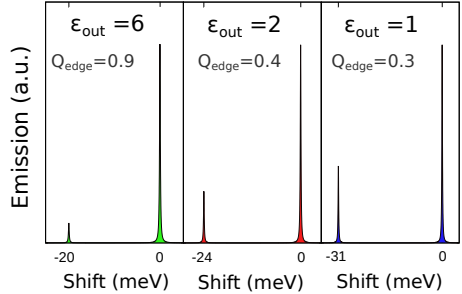}
\caption{Normalized X$^-$ emission as a function of the environment dielectric constant.
With increasing dielectric contrast, the SU peak increases and becomes more redshifted.
For every value of $\epsilon_{out}$, the value of $Q_{edge}$ that maximizes SU transitions
is shown. In all cases, $\epsilon_{in}=6$.}\label{fig4}
\end{figure}
 
We have argued above that strong Coulomb admixture of configurations facilitates the appearance of SU processes.
A distinct feature of colloidal NPLs when compared to epitaxial quantum wells is the presence
of a prounounced dielectric contrast with the organic ligands surrounding the NPL,
which enhances Coulomb interactions by effectively reducing the system dielectric screening.\cite{AchtsteinNL,RajadellPRB,BenchamekhPRB}
To study the influence of this phenomenon over SU transitions, in Figure \ref{fig4}
we compare the trion emission spectrum for different values of the environment
dielectric constant $\epsilon_{out}$, while fixing that of the NPL to the high-frequency
CdSe value, $\epsilon_{in}=6$. For the sake of comparison, the emission spectrum is 
normalized so that the band edge peak has the same intensity in all cases. 
Also, we have selected the value of $Q_{edge}$ that maximizes the relative size of the SU peak in each case.
Because $\epsilon_{out}$ screens the surface charge electrostatic field, 
larger $Q_{edge}$ values are needed when $\epsilon_{out}$ increases.
The figure evidences that lowering $\epsilon_{out}$ increases the SU peak height 
and energetic redshift. % with respect to the band edge transition.  
For typical ligands of CdSe NPLs (e.g. oleic acid), $\epsilon_{out} \sim 2$.\cite{AchtsteinNL,EvenPCCP}
We then conclude that dielectric confinement makes SU processes in colloidal NPLs
more conspicuous.

\subsection{Core/shell NPLs}

\begin{figure}
	\centering
		\includegraphics[width=13cm]{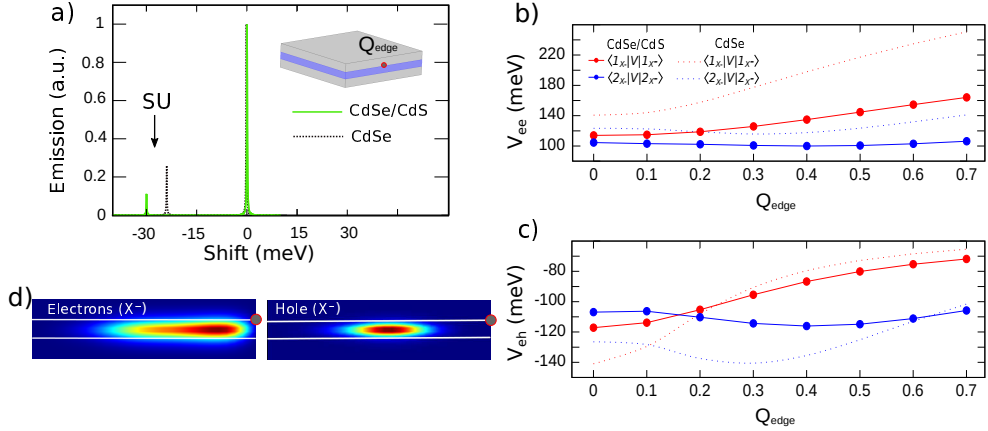}
\caption{(a) Normalized X$^-$ emission spectrum a CdSe/CdS NPL with 6 ML-thick shell (solid line), 
compared to that of a core-only CdSe NPL (dotted line). 
The spectra are centered at the energy of band edge transition.
$Q_{edge}=0.6$ ($0.4$) is used for the CdSe/CdS NPL (core-only NPL), to maximize the relative height of SU lines.
The SU peak for the core/shell system is smaller than for core-only NPLs. 
 %The inset shows the geometry under study.
(b,c) Average Coulomb integrals of $|GS_{X^-}\rangle$ configurations 
$|1_{X^-}\rangle$ and $|2_{X^-}\rangle$: (b) electron-electron repulsion, (c) electron-hole attraction.
Solid (dotted) lines are used for core/shell (core-only) NPLs. The interactions
are weaker in the core/shell structure.
(d) Charge densities of %single electron (left), 
electrons (left) and hole (right) for the trion ground state in the CdSe/CdS NPL at $Q_{edge}=0.6$.  
The electron stays in the vicinity of the core, despite the shallow band offset.}\label{fig5}
\end{figure}

We next consider heterostructured core/shell NPLs. The first case under study are CdSe/CdS NPLs.\cite{TessierNL,AchtsteinACS,PolovitsynCM,LlusarJPCc} 
%These are quasi-type-II structures, with a shallow CB offset which is further reduced by strain.\cite{LlusarJPCc}
%The strong quantum confinement in the vertical direction provides independent electrons with 
%sufficient kinetic energy so as to delocalize all over the structure, while the hole stays in the core.\cite{TessierNL,AchtsteinACS,PolovitsynCM}
%
 The NPLs have the same CdSe core as in the previous section and 6 ML thick CdS shells
on top and bottom  (see inset in Figure \ref{fig5}a).
In general, the behavior of SU replicas is found to be analogous to that of core-only NPLs. 
An off-centered acceptor impurity is needed to yield sizable SU replicas,
with an optimal value of $Q_{edge}$ maximizing the relative size of the SU peak. 

Figure \ref{fig5}a shows the emission spectrum of X$^-$ for the optimal $Q_{edge}$ value,
in CdSe/CdS NPLs (green line) against CdSe core-only NPLs (black, dashed line).
One can see that the SU replica of the CdSe/CdS structure is again significant
 (11\% of the main transition), but less pronounced than in the core-only structure 
 (26\%). 
The smaller SU replica in the core/shell structures is a robust result, which holds
for different shell thickness and surface charge locations. It is a consequence of the
weaker Coulomb interactions. The electron leakage into the CdS shell reduces 
electron-electron repulsions and electron-hole attractions. 
 The quenching of dielectric confinement by the CdS shell, which pushes organic ligands far from
the core, further contributes to the weakening.
This observation is reflected by Figs.~\ref{fig5}b and \ref{fig5}c, 
which show that Coulomb interactions (especially $V_{ee}$) are weakened in core/shell NPLs 
(solid lines) as compared to core-only NPLs (dotted lines). 
Configuration $|2_{X^-}\rangle$ is then less stabilized with respect to $|1_{X^-}\rangle$,
which implies smaller $|c_2|$ coefficient in the CI expansion.

Figure \ref{fig5}d compares the charge density of the two electrons (left) 
and hole (right) in $|GS_{X^-}\rangle$.
%The independent electron largely delocalizes over the CdS shell, but 
%for the trion electrons stay in the vicinity of the core %, %rather than delocalizing all over the structure, 
 The trion electrons are found to stay in the vicinity of the core, rather than delocalizing all over the structure, 
 to benefit from interaction with the hole. 
 This is consistent with the observed behavior of CdSe/CdS NPLs being similar to that
 of core-only structures, albeit with weakened Coulomb interactions due to the lessened confinement.

\begin{figure}
\centering
\includegraphics[width=6cm]{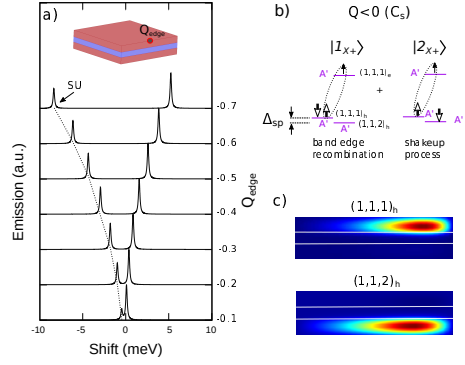}
\caption{(a) Normalized X$^+$ emission spectrum in a CdSe/CdTe NPL
with 6 ML thick shells, as a function of the lateral charge strength.
Dotted line is a guide to the eye. SU peaks and fundamental transition
have comparable intensities.
(b) Two main $|GS_{X^+}\rangle$ configurations in the CI expansion
in the presence of a charge. The weight of $|2_{X^+}\rangle$ is comparable
to that of $|1_{X^+}\rangle$ in this system, which explains the high SU peaks in (a).
(c) Wave function of $(1,1,1)_h$ and $(1,1,2)_h$ hole orbitals under $Q_{edge}=-0.5$.
The states have the same symmetry but localize on opposite sides of the core
to stay orthogonal.}\label{fig6}
\end{figure}

Understanding the conditions which promote SU processes allows us to devise
structures where their impact would be maximal. In Fig.~\ref{fig6} we consider 
a core/shell NPL with the same dimensions as before, but CdSe/CdTe composition.
The NPL is chosen to be charged with a positive trion (X$^+$).
Because of the type-II band alignment, the electron stays in the CdSe
core and the holes in the CdTe region, as observed in related core/crown structures.\cite{AntanovichNS,KelestemurJPCc}
In the absence of external charges, the two first hole orbitals are $(1,1,1)_h$ and $(1,1,2)_h$, 
i.e. the symmetric ($A_{1g}$) and antisymmetric ($A_{1u}$) solutions of the double well potential, respectively,
which are almost degenerate because tunneling across the core is negligible
(i.e. $\Delta_{sp} \rightarrow 0$). Switching on a negative surface charge, 
$Q_{edge} < 0$, lifts the inversion symmetry so that both orbitals acquire $A'$ symmetry
and can be Coulomb coupled. 
The admixture between configurations $|1_{X^+}\rangle$ and $|2_{X^+}\rangle$, 
depicted in Fig.~\ref{fig6}b, is then very strong.
 In the presence of the charge, the two hole orbitals tend to localize 
on opposite shell sides to remain orthogonal, as shown in Fig.~\ref{fig6}c. 
This implies that configuration $|1_{X^+}\rangle$, which has two holes
in the same orbital, has much stronger repulsion than configuration $|2_{X^+}\rangle$, 
which distributes the two electrons on opposite sides of the core. This makes
$\langle 1_{X^+} | V_{hh} | 1_{X^+} \rangle \gg \langle 2_{X^+} | V_{hh} | 2_{X^+}\rangle$. 
Altogether, the small $\Delta_{sp}$ value and the large difference in hole-hole repulsion
explain the strong admixture between configurations $|1_{X^+}\rangle$ and $|2_{X^+}\rangle$.
As shown in Fig.~\ref{fig6}a, this gives rise to SU peaks whose magnitude is almost
as large as that of the fundamental transition ($72\%$ for $Q_{edge}=-0.5$).

\section{Discussion}

Our simulations show that SU processes can be expected for trions in core-only 
and core/shell NPLs, if off-centered impurities are present.  %
We discuss here the potential relationship of this finding with experiments
and practical implications.

\subsection{Relationship with experiments}

In core-only CdSe NPLs, the low temperature photoluminescence is thought to arise 
from subpopulations of excitons and negative trions.\cite{ShornikovaNL,YuAMI,AntolinezNL2} 
Very recently, Antolinez and co-workers have reported that the X$^-$ 
emission shows a distinct peak or a shoulder (depending on the film thickness)
redshifted from the trion band edge transition. The redshift is $\sim 19$ meV 
and the relative height $15-25\%$ that of the main peak.\cite{AntolinezNL2} 
They speculated that the origin could be a SU process of the kind we study.
Our calculations support the feasibility of this interpretation.
Figure \ref{fig1}a shows excellent agreement with the experimental measurements,
both in energy and relative intensity of the SU peak, 
assuming a lateral charge with $Q_{edge}=0.3-0.4$,
which gives a redshift of $19-25$ meV and a relative height of $15-23$ \%.

The presence of acceptor impurities in CdSe NPLs likely originates when
the hole of a photoexcited electron-hole pair is trapped by a surface
defect. The next electron-hole pair generated in the NPL joins the
residual electron to form X$^-$, while the trapped hole exerts 
a screened electrostatic potential.\cite{RabouwNL,FengNL,CalifanoNL}
The coexistence of X$^-$ and trapped surface charges in CdSe NPLs 
is backed up by studies reporting correlation between surface-to-volume ratio, 
laser irradiation time and trion emission intensity.\cite{YuAMI}
A plausible location for surface charges are the
lateral sidewalls of the NPL (as in Fig.~\ref{fig1}b).
This possibility is suggested by studies showing that Z-type 
ligand desorption --and hence surface traps-- in CdSe NPLs is more 
frequent on these facets,\cite{LeemansJPCL}
and by the fact that CdSe/CdS core/crown NPLs generally improve the
photoluminescence quantum yield as compared to core-only structures,
despite having larger surfaces on top and bottom.\cite{TessierNL2}
Because off-centered charges are needed to originate SU peaks, 
lateral charges are candidates to trigger such processes.\\

In core/shell CdSe/CdS NPLs, SU processes have been also proposed 
as the origin of multi-peaked fluorescence emission 
--and hence broadened line width--.\cite{AntolinezNL}
Our simulations in Fig.~\ref{fig5}a confirm one can indeed expect 
a sizable SU peak in such structures. %again in the presence of lateral $Q_{edge}$ charges.
We note that earlier experimental studies had so far interpreted the
line width broadening as a result of either SU processes\cite{AntolinezNL}
or of surface defects.\cite{TessierNL} By showing that the second effect
is a prerequesite for the first one, our study helps to reconcile both
interpretations.
Nonetheless, two remarkable disagreements are observed between our simulations
and Ref.~\cite{AntolinezNL} measurements. 
First, the experiments show from 2 to 4 emission peaks, 
which are interpreted as the X$^-$ fundamental transition plus up to three redshifted, SU peaks.  
In our calculations, however, we fail to see more than one significant SU replica.
Second, the highest-energy peak in the experiment is never the brightest one. 
This is inconsistent with our results and with earlier studies on epitaxial
quantum wells and dots, where the higher-energy peak corresponds to the fundamental
transition, which is the most likely recombination channel.\cite{NashPRL,FinkelsteinPRB,BryjaPRB,DzyubenkoPRB,PaskovPE}

Tentatively, one may suspect that a large number of SU peaks in core/shell CdSe/CdS NPLs 
could be connected with the thick CdS shell (12 ML in Ref.~\cite{AntolinezNL}), which
makes surface defects more likely than in core-only structures. 
A significant presence of defects in these structures has been hinted 
by studies showing that the long radiative lifetime is not due to electron delocalization 
but to the influence of impurities.\cite{AchtsteinACS}
However, Coulomb interactions are weaker than in core-only structures
(Fig.~\ref{fig5}c,d), where only one SU peak has been measured.\cite{AntolinezNL2} 
It is then not surprising that, despite investigating different charge locations
(Figs. S3, S6 and S7 in SI), conduction band-offset values (Fig. S4) and
shell thicknesses (Fig.S5), we see at most one significant SU satellite.
 
Regarding the relative intensity of the peaks,
%the highest energy peak corresponds to fully radiative emission of X$^-$, while
%lower energy peaks correspond to partial non-radiative energy transfer to the remaining electron.
 as mentioned in the previous section, the highest-energy one (fundamental transition) 
 is proportional to the weight of configuration $|1_{X^-}\rangle$ in the CI expansion, $|c_1|^2$, 
while subsequent (SU) peaks would be proportional to $|c_2|^2$, $|c_3|^2$, \ldots
Configuration $|1_{X^-}\rangle$ (all carriers in the ground orbital, Fig.~\ref{fig2}c) 
is nodeless and hence naturally expected to be the dominant one, 
so the highest-energy peak is also the brightest one.
We have not observed SU peaks exceeding the fundamental transition height
 despite considering different charge locations and shell thicknesses (see SI). 
 Even in CdSe/CdTe NPLs, which constitute a limit case, 
 SU peaks never exceed the height of the main transition, see Fig.~\ref{fig6}a.

\begin{figure}
\centering
\includegraphics[width=6cm]{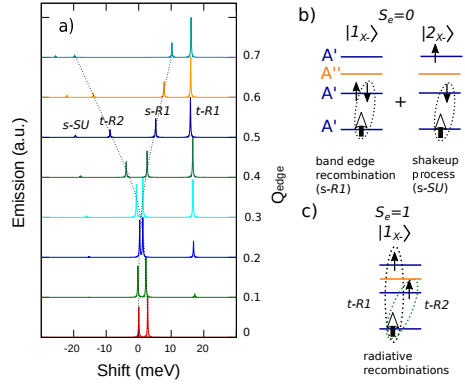}
\caption{(a) Normalized X$^-$ emission spectrum in a CdSe/CdS NPL,
 as a function of the lateral charge strength.
An electron spin relaxation bottleneck is imposed, so that emission comes from 
the lowest singlet ($S_e=0$, ground state) and triplet ($S_e=1$) states.
Dotted lines are guides to the eyes.
(b-c) Sketches showing the relevant electron-hole recombination channels
of singlet and triplet states. (b) The singlet can give rise to one fully radiative 
($s$-$R1$) plus one SU transition ($s$-$SU$). 
(c) The triplet can give rise to two fully radiative transitions, $t$-$R1$
and $t$-$R2$.}\label{fig7}
\end{figure}

As an alternative interpretation for the experiments, a multi-peaked emission spectrum could 
result from stacking of colloidal NPLs,\cite{DirollNL} which leads to electronic 
coupling through dielectric confinement.\cite{MovillaJPCL}
However, the time-dependent spectral shifts observed by Antolinez \emph{et al.} 
suggest that all peaks arise from the same NPL, and significant stacking was not expected
in the experiment samples.\cite{AntolinezNL}
We thus propose a different interpretation. Namely, simultaneous emission from the X$^-$ 
ground state, with singlet electron spin ($S_e=0$), and a metastable excited state with triplet
electron spin ($S_e=1$). The decay from the triplet to the singlet state is slowed down 
by spin selection rules, as phonons are spinless. This could allow simultaneous occupation
of the two states even if the energy splitting exceeds thermal energy. %, as already noted in 
%the photoluminescence of epitaxial quantum wells.\cite{BryjaPRB,ShieldsPRB}
 
To illustrate this point, in Figure \ref{fig7}a we show the calculated emission of X$^-$ 
assuming equipopulation of $S_e=0$ and $S_e=1$ trion states. One can see that the number 
of sizable peaks in the spectrum ranges from two to four, depending on the 
strength of surface charge, $Q_{edge}$.
%
%shows the calculated emission spectrum of X$^-$ assuming equipopulation of 
%the lowest $S_e=0$ and $S_e=1$ states, as a function of the edge charge strength.
%One can see that two to four emission peaks may arise, depending on the value of $Q_{edge}$.
 The origin of these peaks is summarized in the sketches of Fig.~\ref{fig7}b and \ref{fig7}c.
The singlet (Fig.~\ref{fig7}b) can give rise to a fully radiative transition ($s$-$R1$) 
and a SU transition ($s$-$SU$), as described in the previous sections. 
In turn, the triplet (Fig.~\ref{fig7}c) can give rise to two fully radiative transitions 
($t$-$R1$ and $t$-$R2$), depending on which electron recombines with the hole.
$t$-$R2$ is readily visible at $Q_{edge}=0$, but $t$-$R1$ requires recombining the hole 
with an excited electron, a process which is again activated when the surface charge 
lifts symmetry restrictions.
However, unlike in SU processes, the two triplet transitions come from the main configuration 
of the trion CI expansion. Therefore, their intensity can be comparable to that of the band
edge transition, $s$-$SU$, even if Coulomb admixture is weak. % (see e.g. $Q_{edge}=0.3-0.5$).
The triplet transitions are built on both sides of $s$-$SU$, with inter-peak energy 
splittings up to few tens of meV.
The relative sizes of the peaks will be further modulated in realistic situations
by a finite triplet-singlet decay rate. This relaxation channel would possibly reduce the relative 
population of $S_e=1$, and hence the intensity of $t$-$R1$.  

Altogether, the number of peaks, the magnitude of the energy splitting between the peaks
and the flexible intensities provide a framework to explain the 
multi-peaked photoluminescence of Ref.~\cite{AntolinezNL}.
Several other aspects of this proposal are consistent with the experiments.
For example, because all peaks in Fig.~\ref{fig7}a arise from the same NPL, 
they will experience simultaneous spectral shifts when surface impurities migrate.\cite{AntolinezNL} 
Also, the hot trion emission is expected to vanish when the impurities are removed, 
as $t$-$R1$ becomes deactivated and $t$-$R1$ almost merges with the singlet emission, $s$-$R1$,
see Fig.~\ref{fig7}a for $Q_{edge}=0$. This fits the transition from asymmetric to symmetric band shape
as temperature increases.\cite{TessierNL}

The fact that triplet emission is observed in CdSe/CdS NPLs, but not in CdSe ones,
may be explained from the strong spin-spin interaction of resident carriers and 
surface dangling bonds in the latter case,\cite{ShornikovaNN}
which should speed up spin relaxation through flip/flop processes.
This mechanism is expected to be inhibited in core/shell structures, because
X$^-$ carriers stay far from the surface, as shown in Fig.~\ref{fig5}d.
On the other hand, the triplet trion is expected to have fine structure
through electron-hole exchange interaction\cite{WarePRL}, 
which may not fit the mono-exponential photoluminescence decay reported in Ref.~\cite{AntolinezNL}. 
Further experiments are needed, e.g. on polarisation of the different peaks under external fields\cite{ShornikovaNL,JovanovPRB},
to confirm the different spin of the emissive states in CdSe/CdS NPLs. 

The observation of metastable triplet trion photoluminescence has been previously reported 
in epitaxial quantum wells\cite{BryjaPRB,ShieldsPRB} and dots\cite{JovanovPRB},
and more recently in transition metal chalcogenide monolayers.\cite{VaclavkovaNT}% tmp ***
To our knowledge, however, its presence in colloidal nanostructures has not been confirmed.

\subsection{Control of SU processes}

Inasmuch as SU processes can be responsible for the line width broadening NPLs,
their supression is desirable to improve color purity in optical applications.
It has been suggested that this job could be achieved by increasing quantum confinement,
reducing either lateral dimensions or shell thickness --the latter would favor 
electrostatic confinement.\cite{AntolinezNL} 
Both strategies have the drawback of introducing size dispersion in ensemble
luminescence. 
From our theoretical analysis, we confirm that reducing Coulomb admixture 
would minimize SU processes, but this can be achieved by weakening Coulomb interactions 
instead of increasing quantum confinement. For example, reducing dielectric 
confinement or using thinner cores to enhance the quasi-type-II character 
should contribute to this goal.  Obviously, this approach would have the 
drawback of reducing the band edge recombination rate as well.

Alternatively, since our study shows that impurities are ultimately responsible
for SU processes, experimental routes to suppress SU processes could be directed 
to control of traps. Appropriate choice of surface ligands\cite{LeemansJPCL},  electrochemical 
potentials\cite{GallandNAT} and interface alloying\cite{RossinelliCM,KelestemurACS} 
could contribute to this end. 

Because we find surface charges on lateral sidewalls particularly suited to induce SU processes,
the growth of core/crown heterostructures is expected to reduce their influence by 
keeping the outer rim away from the photogenerated carriers. 
This suggestion seems to agree with experimental observations by Kelestemur and co-workers, 
indicating that core/crown/shell CdSe/CdS NPLs have more symmetric emission behavior than core/shell 
ones at cryogenic temperatures,\cite{KelestemurAFM}
This can be understood as a consequence of the suppression of SU processes in 
the low-energy tail of the emission band. 
It is also consistent with recent single-particle studies showing that the 
line width in CdTe/CdSe core/crown NPLs is set by LO phonon replica, 
rather than SU ones.\cite{SteinmetzJPCc}
%However, the presence of defects on the core/crown interface or cracks in the
%CdS crown can remain an issue, as the line width of CdSe/CdS core/crowns NPLs 
%is also wider than in core-only NPLs.\cite{TessierNL2,KelestemurAFM}

Should the role of metastable triplet states be confirmed in CdSe/CdS NPLs, 
strategies to control the line width should rather focus on enhancing the interaction of
confined carriers with surface spins\cite{ShornikovaNN} or intrinsic spin-orbit interaction\cite{TadjinePRB}, 
to shorten their lifetime.
Replacing trion by neutral exciton emission through thermal dissociation,\cite{AyariNS} 
is yet another possibility to avoid SU and high spin peaks.

Thus, our calculations propose a wealth of experiments targeted at material design to 
tune quantum and dielectric confinement, and exciton-surface/interface interactions, 
and set suitable temperature ranges to control SU/triplet emission.

\section{Conclusions}

We have shown that SU processes in colloidal NPLs are enabled by severe Coulomb admixture 
--which results from strong Coulomb interactions and weak lateral confinement-- 
and the presence of off-centered electrostatic traps, which suppress the %otherwise natural 
 protection against Auger processes provided by symmetry conservation. 
Surface charges on lateral sidewalls seem particularly efficient to this end. 

Under typical experimental conditions, core-only and core/shell NPLs are susceptible of showing a SU peak
with oscillator strength $0.1$-$0.3$ times that of the band edge transition.
This is at least one order magnitude larger than in epitaxial quantum wells.
The SU peak is redshifted from the band edge peak by up to few tens of meV,
thus providing a source of line width broadening.
 
These results are in excellent agreement with recent experimental findings in CdSe NPLs\cite{AntolinezNL2}
in terms of number of emission peaks, energy splitting and relative intensity,
but only partially so with those of core/shell CdSe/CdS NPLs.\cite{AntolinezNL}
Experiments in the latter structure are however in line with an alternative interpretation 
involving simultaneous participation from trion singlet and metastable triplet states.

Strategies to narrow the line width of NPLs through suppression of SU processes should
aim at controlling electrostatic impurities or Coulomb admixture.

\section{Methods}

Calculations are carried within k$\cdot$p-continuum elastic theory framework.
Independent electron and hole states are calculated with single-band Hamiltonians
including strain and self-energy potential terms. 
Model details and material parameters are given in Ref.~\cite{LlusarJPCc}.
Point charge electrostatic potentials and Coulomb integrals for 
CI matrix elements, including dielectric mismatch effects, 
are calculated solving Poisson Equation with Comsol $4.2$.
The CI basis set is formed by all possible combinations of the first
22 single-electron and 22 single-hole spin-orbitals. For X$^-$,
these are combined to form configurations $|m_{X^-}\rangle$ 
as the Hatree product of one hole spin-orbital with a two-electron 
Slater determinant.

\begin{acknowledgement}
The authors acknowledge support from MICINN project CTQ2017-83781-P.
We are grateful to I. Moreels, A. Achtstein and F. Rabouw for useful discussions.
\end{acknowledgement}

{\bf Supporting Information Available:} 

Additional calculations on the influence of trion charge, surface charge
location and conduction band offset over the formation of SU processes are
provided.

\bibliography{shakebib}

\end{document}

%\newpage

\newpage

\section{Additional calculations}

We present here additional calculations for further understanding of SU processes. 
%We start by some extra investigations we made on negative charged trion. First, we display some additional impurity locations from main text, such as: (a) located at corner or (b) placed at an hybrid location: edge-top and corner-top. Therefore, we will see why the edge-located impurity was the chosen one. Next, we will show the effect of using a lower CBO\cite{SteinerNL} than that reported in the main text for a CdSe/CdS system. Right after, it is demonstrated a correspondence about charge location between our main text system of CdSe/CdS (6ML shell) and the same system with a 12ML shell and their implications. Also, we will see the effect of introducing multiple impurities onto the 12ML shell system of CdSe/CdS. Finally, we will show briefly how behaves $X^{+}$ in Core-only (CdSe). We shall see similar properties that those seen on $X^{-}$. 

\subsection{Convergence of CI calculations}

Configuration Interaction (CI) calculations on the basis of independent particle (or Hartree-Fock) orbitals provide an excellent description of repulsions in few- and many-fermion systems\cite{JacakSpringer,RontaniJCP}. However, large basis sets are needed to describe strong attractions,\cite{RontaniJPB,ShumwayPRB} which are certainly present in colloidal NPLs\cite{RajadellPRB} and are involved in a correct description of SU processes.

\begin{figure}
	\centering
		\includegraphics[width=8cm]{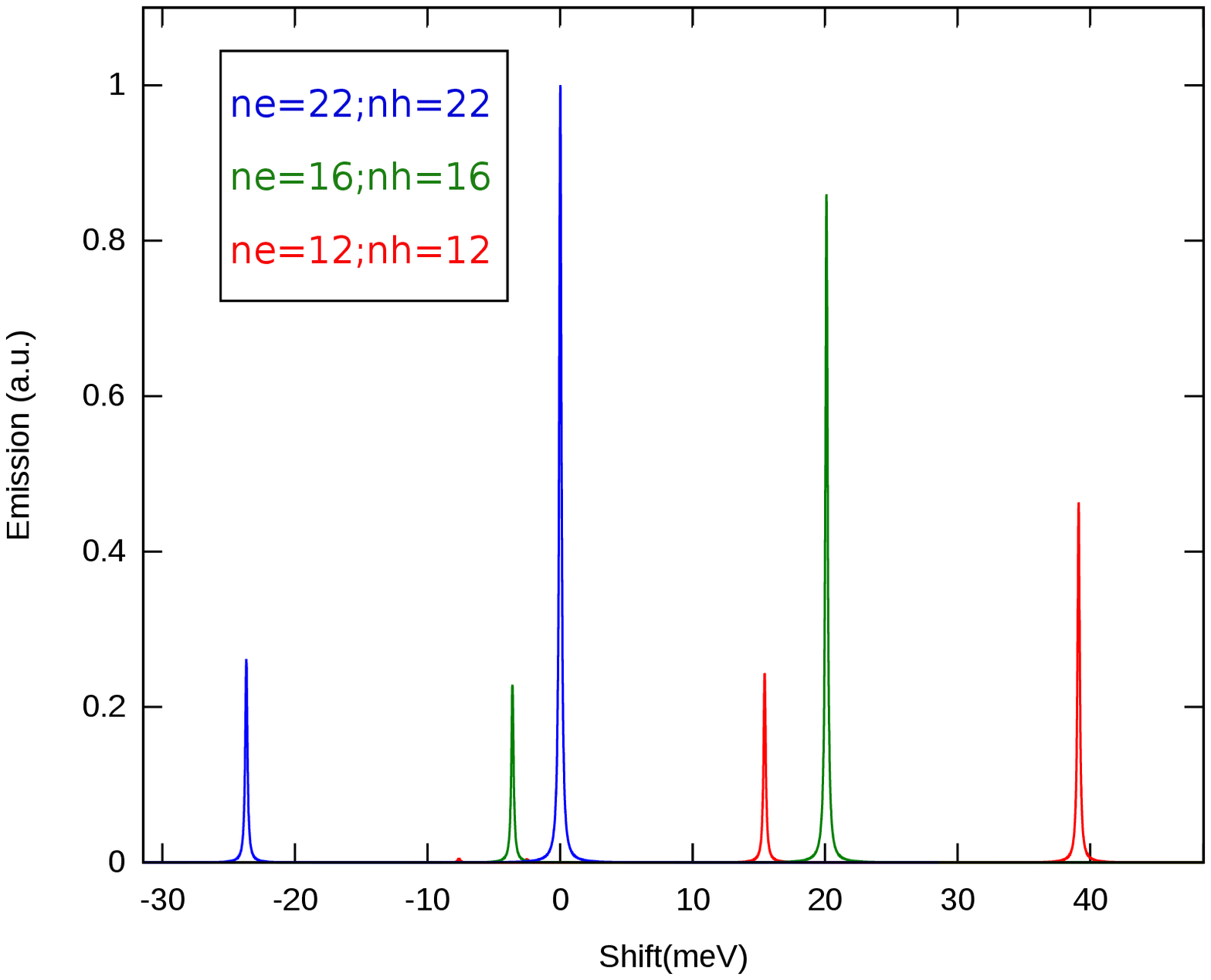}
		\caption{X$^-$ emission spectrum for $Q_{edge}=0.4$ (see main text). Zero energy is set for the fundamental transition with $ne=nh=22$.
$ne$ and $nh$ are the number of single-electron and single-hole spin-orbitals, respectively, used to build the CI basis sets.}
	%{\bf La figura hauria de dir S1, no 1. Mira .tex supp info de article strain}  } % tmp ***
\label{figS}
\end{figure}
 
 In Fig.~\ref{figS} we compare the X$^-$ emission spectrum calculated for CdSe NPLs --same dimensions as in main text-- 
using different basis sizes.  The basis is formed by all possible combinations of the first $ne$ ($nh$) independent particle 
 spin-orbital states of electrons (holes). 
 With increasing basis dimensions, the band edge transition peak redshiftes and gains intensity, which reveals an improved
 description of electron-hole correlation. The intensity of the SU peak height and its redshift with respect to the 
 band edge transition are however less sensitive to the basis dimensions. %because they rely on the CI configurations $|2\rangle$ (see main text), 
% which involves low-energy orbitals, already contained within the smallest basis set we use.
 %
 It follows from the figure that quantitative assessment on the ratio of fundamental vs SU peak heights requires 
 large basis sets. In the main text we use $ne=nh=22$. By comparing with smaller values of $ne/nh$ in the figure,
 it is clear that for this value --which involves very time-consuming computations-- the ratio is reaching saturation.
 This validates the order of ratios provided in the main text.
 For the calculations in this Supporting Information, however, we may resort to $ne=nh=12$, 
 which overestimates the relative height of SU peaks, but suffices to provide qualitative assessment.

\subsection{Positive trion behaviour} \label{positive_trion}

\begin{figure}
	\centering
		\includegraphics[width=8cm]{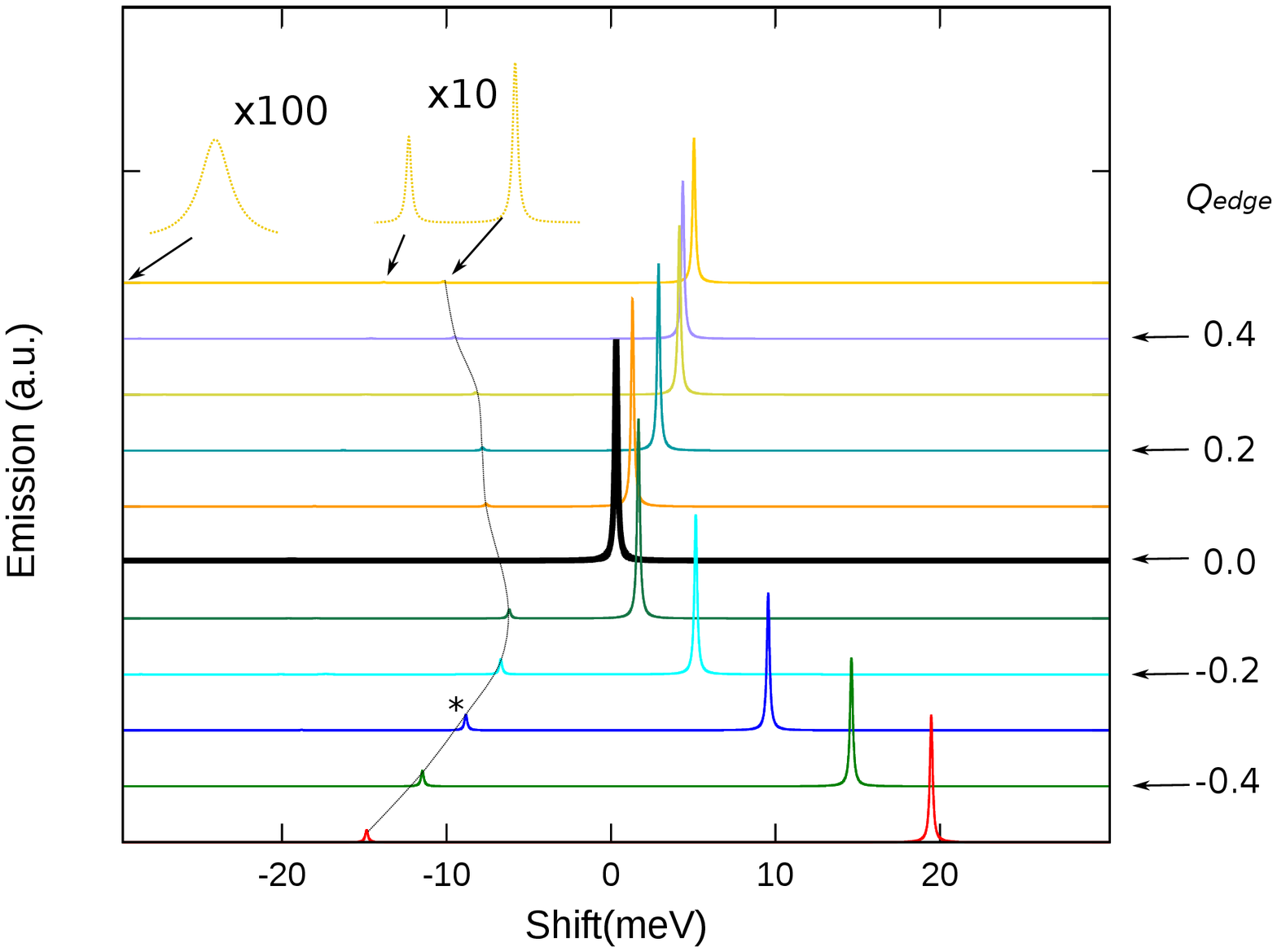}
\caption{$X^{+}$ normalized spectrum emission for different charge intensities. The arrows are pointing to SU satellites (dotted lines are guides to the eyes). The highest SU peak ($Q_{edge}=-0.3$) is marked with a star. The origin of energies is set at the band edge recombination peak. The insets correspond to $Q_{edge}=0.5$ amplified SU peaks.}
\label{figSXXX3}
\end{figure}

In the main text, we have mostly considered the case of negative trions.
We show here that the same behavior holds for positive ones.
To illustrate this point, we choose the case of the core-only NPL with an edge charge,
equivalent to Fig.1d of the main text.
Figure \ref{figSXXX3} shows that the presence of SU peaks in the emission 
spectrum is again strongly dependent on the value of the surface charge.
For $Q_{edge}=0$, no SU peak is observed. For repulsive ($Q_{edge}>0$) charges,
SU are formed but very small in magnitude. The highest SU peaks are formed
for weakly bound donor charges ($Q_{edge}<0$), which attract the holes of X$^+$,
marked with a star in the figure. As in the X$^-$ case, if the attractive charge
further increases it starts dissociating the trion. 
Consequently, SU peaks are quenched again.
Notice however energy splittings for X$^+$ (Fig.~\ref{figSXXX3}) 
are smaller than for X$^-$ (Fig.~1d in the main text).
This is expected from the heavier masses of holes.

\subsection{Effect of charge impurity location}

\begin{figure}
	\centering
		\includegraphics[width=16cm]{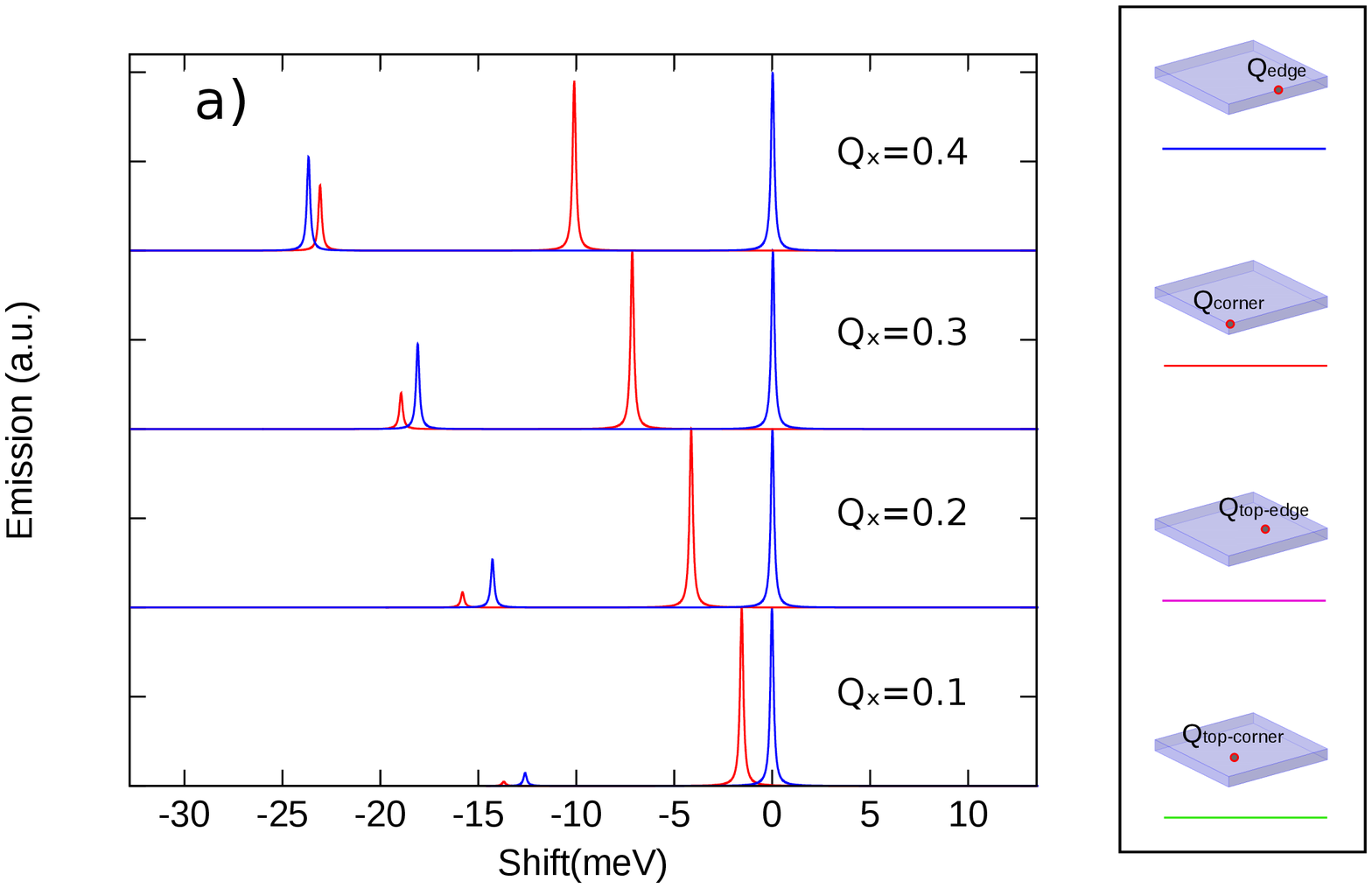}
\caption{X$^-$ emission spectra for different locations of surface charges. 
The spectra are normalized with respect to the energy and intensity of the $Q_{edge}$ fundamental transition.
(a) Edge-located vs. corner-located impurity. Blue and red lines stand for edge and corner, respectively. 
(b) Edge-located vs. edge-top-located vs. corner-top-located. Blue, green and pink lines stand for edge, 
top-corner and top-edge, respectively. $ne=nh=12$.}
\label{figSX}
\end{figure}

In the main text we present the representative cases of a surface charge centered on top of the NPL ($Q_{top}$), 
and centered and that of a charge on the edge of lateral sidewall ($Q_{edge}$). In Figure ~\ref{figSX} we compare with
different locations. One can see that the effect of a charge located in the corner, red line in Fig.~\ref{figSX}a, 
provides similar SU peaks to that of the edge charge, blue line in the figure, both in energy and intensity.
We recall that these traps seem to be particularly likely according to recent studies on ligand desorption.\cite{LeemansJPCL,DrijversCM}
Off-centered charges on top and bottom surfaces are studied in Fig.~\ref{figSX}b. They give rise to SU peaks
of similar height to that of $Q_{edge}$, although they reach the optimal charge value sooner than $Q_{edge}$
($Q_{top-edge} \sim Q_{top-corner} \approx 0.2$ versus $Q_{edge}=0.4$), 
because they lie closer to the center of the NPL, where photogenerated carriers tend to localize.

%In order to understand why SU peaks are greater ones of the others, we may ask ourselves: (a) how far is the impurity located from the trion charge density and (b) how intense is this impurity charge. On one hand, when comparing between edge and corner locations, Fig.S\ref{figSX}a, we observe that SU peaks where the impurity is edge-located, reach faster its optimal charge than that located on corner. On the other hand, when comparing edge vs. top-corner and top-edge locations, Fig.S\ref{figSX}b, top-corner and top-edge locations reach faster their optimal charge than the edge-located one does. These observations are in agreement with (a) and (b). Further, having in mind the emission spectrum of top and edge in the main text, (a) and (b) also apply in these cases. 

\subsection{Effect of conduction band offset in CdSe/CdS NPLs} \label{red_cbo}

\begin{figure}
	\centering
		\includegraphics[width=16cm]{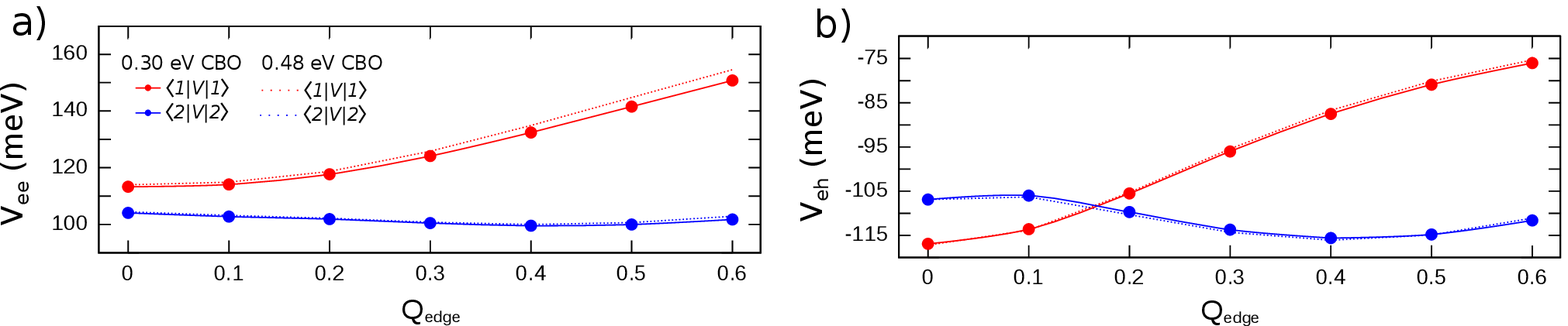}
	\caption{(a,b) Average Coulomb integrals for $Q_{edge}=0.6$: (a) electron-electron repulsion and (b) electron-hole attraction for every CBO. $ne=nh=22$}
\label{figSXX}
\end{figure}

The value of the CdSe/CdS conduction band offset (CBO) has been a subject of debate in nanocrystal heterostructures.\cite{AIPZunger,SteinerNL,LlusarJPCc}. We used, along our main text, an upper-bound unstrained value of $0.48$ eV\cite{AIPZunger}, which is partly reduced by compressive strain in the core.\cite{LlusarJPCc} Here we explore the scenario where we use a lower-bound\cite{SteinerNL} value as well, to see the possible effect of enhancing electron delocalization over the CdS shell. Figure \ref{figSXX} compares the two cases.
Lowering the CBO gives rise to slightly weaker electron-electron repulsion ($V_{ee}$) and electron-hole attraction ($V_{eh}$),
however the differences are very small. One can then expect similar role of SU processes as in the main text.
%{\bf Poses densitats en la fig pero no es diu res al respecte? Millor no dir "normal" i "reduced" sino 0.48 i 0.30. } % tmp ***

\newpage

\subsection{Effect of shell thickness in CdSe/CdS NPLs} \label{correspondence}

Along the main text, core/shell NPLs under study had a shell thickness of 6ML on each side of the core. 
The experiments of Antolinez et al.\cite{AntolinezNL} however used thicker shells (12 ML). 
In this section we compare qualitatively the response in the two cases using a moderate basis set ($ne=nh=12$),
which permits addressing the experimental dimensions without the computational burden of the 
large basis set (for 12 ML thickness, the extended CI computation is beyond our current resources).

\begin{figure}
	\centering
		\includegraphics[width=16cm]{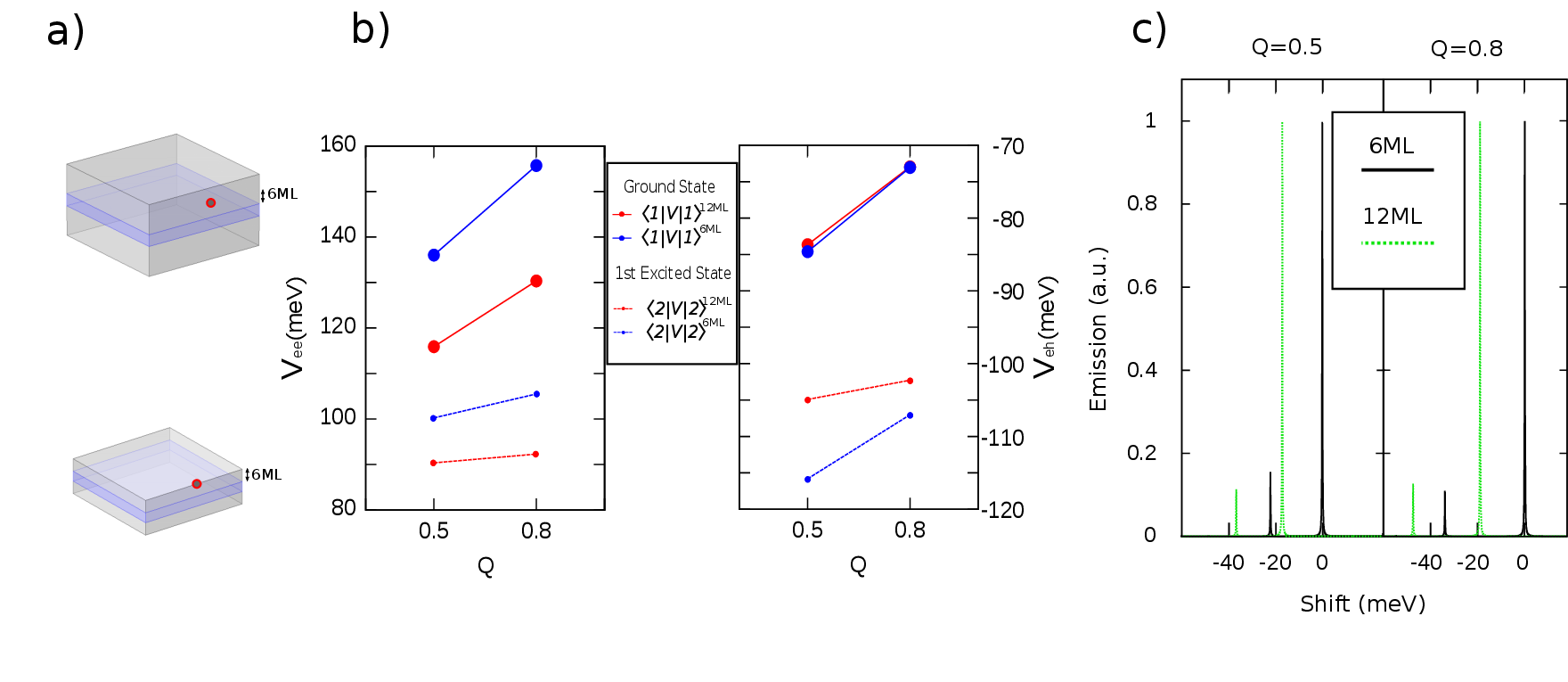}
\caption{(a) Sketch of the NPLs we are comparing: 12ML shell (top) and 6ML (bottom). 
The charge is located at same coordinates. 
(b,c) Coulomb interactions: (b) repulsions e-e and (c) attractions e-h for $Q=0.5$ and $Q=0.8$. 
(d,e) Normalized emission spectra of 6ML vs 12ML: (d) $Q=0.5$ and (e) $Q=0.8$; 
$Q=0$ is centred at band edge recombination energy for 6ML in both cases. $ne=nh=12$}
\label{figSXXX}
\end{figure}

If we focus on the charge location in both systems, Fig.~\ref{figSXXX}a, one may expect similar behaviour. 
The main difference, as can be seen in Fig.~\ref{figSXXX}b (left panel) occurs for repulsive electron-electron interactions,
which are slightly weaker for thick shells. This is a consequence of the larger electron delocalization,
which translates into smaller $|c_2|$ coefficients in the CI expansion (see main text) and hence slightly smaller SU satellite,
as observed in Fig.~\ref{figSXXX}c.

%However, there are subtle differences due to shell thickness. On one hand we may see from Fig.S\ref{figSXXX}b (right panel), electron-hole attraction interactions are quite similar for both systems in terms of configuration $\ket{1}$. Although, they are not the same system. On the other hand, electron-electron repulsion interaction makes the difference, see Fig.S\ref{figSXXX}b (left panel). First, there are less confinement effects in 12 ML than in 6ML (electrons are allowed to delocalize easier, since they feel themselves less squeezed because of shell thickness). Second, as a consequence of confinement effects, dielectric effects are decreased\cite{RajadellPRB}. Then, with a 12ML shell system, we are able to provoke less electron-electron repulsion. Subsequently, we could obtain sizeable SU peaks as we reduce the electron-electron repulsion (check following section). So, because of small variations on electron-electron repulsion and electron-hole attraction on CdSe/12CdS, we are able to reverse the ratio $I_{b-e}/I_{su}$ ($I_{b-e}$ is the band edge recombination intensity and $I_{su}$ is the SU process intensity) when comparing 6ML and 12ML shell thickness. 

\subsection{Effect of inserting multiple impurities in CdSe/CdS} \label{impurity_effect}

\begin{figure}
	\centering
		\includegraphics[width=16.5cm]{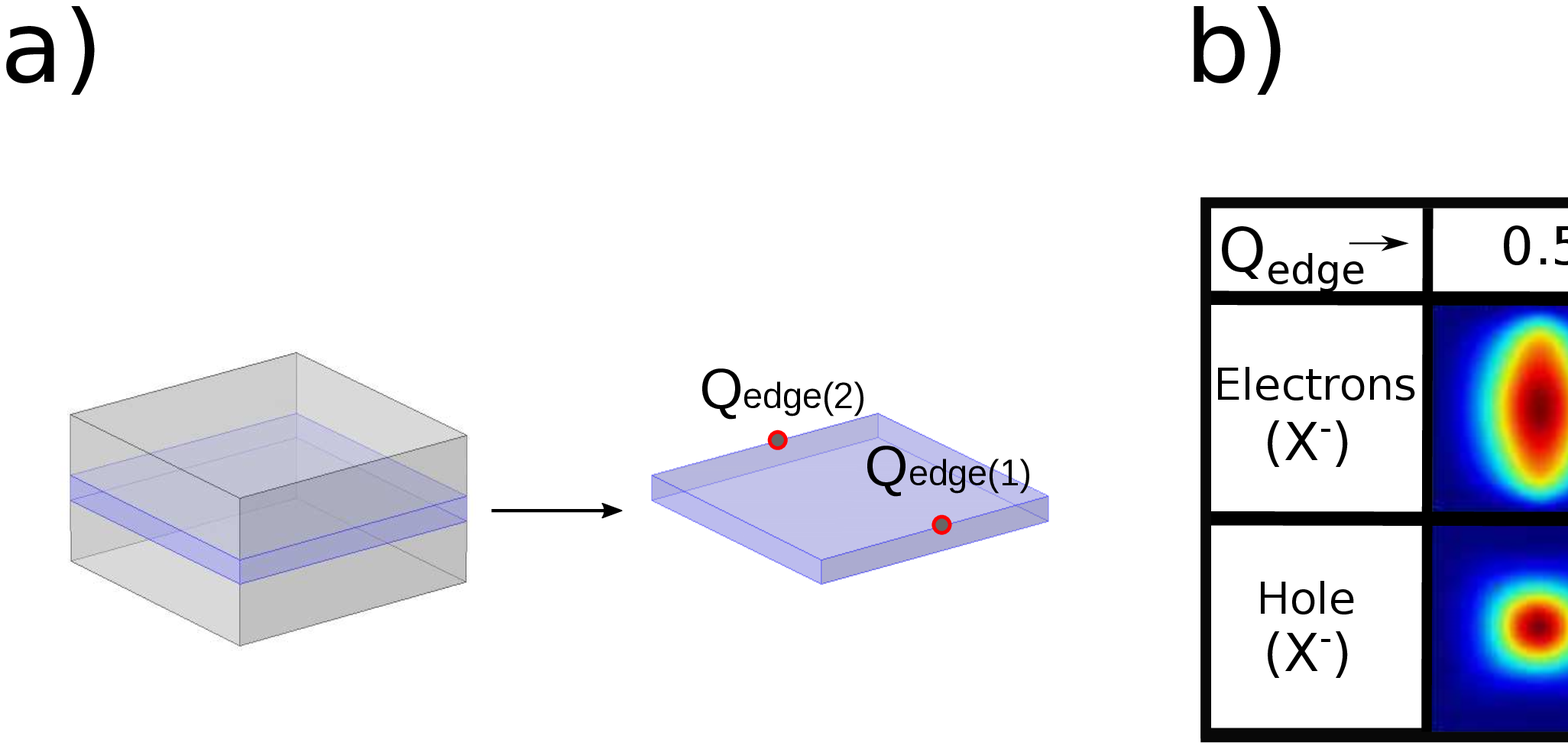}
\caption{(a) Schematic of a CdSe/CdS NPL with 2 charges on edges intersecting interface and sidewall facet. 
$Q_{edge(1)}=Q_{edge(2)}=Q_{edge}$. The NPL shell is 12 ML thick. 
(b) In-plane electrons and hole charge densities for the X$^{-}$ singlet ($S=0$) ground state at $Q=0.5$ and $Q=1.0$; 
(c) Normalized emission spectra at $Q=0.5$ (left) and $Q=1.0$ (right). $ne=nh=12$.}
\label{figSXXX1}
\end{figure}

We consider here the possibility that two surface traps, instead of one, 
are acting as electrostatic impurities in CdSe/CdS NPLs.
Since there is a general preference of forming defects in the heterostructure interfaces 
-- because of lattice mismatch\cite{LlusarJPCc,LiACS} -- and on lateral facets 
-- where ligand desorption is more likely to happen\cite{LeemansJPCL}--, we choose the charges
to be located as shown in Fig.~\ref{figSXXX1}a.
The presence of two charges, combined with the weak in-plane confinement, easily dissociates the trion 
by driving one electron to each surface impurity. This can be seen in the charge densities of Fig.~\ref{figSXXX1}b. The number of visible SU peaks, however, remains one (see Fig.~\ref{figSXXX1}c).
%{\bf Jo no posaria coses en la figura que no es comentaven en el text. P.ex. no necessitem funcions
%d'ona pq ja es veu en el trio que els electrons es separen a pesar del forat. Tanmateix, no necessitem detalls de
%densitat del triplet si no diem res.} % tmp ***
In the case of strong surface charges ($Q=1.0$), the trion triplet (discussed in the main text) becomes so close in 
energy to the singlet ground state that it shows up in the spectrum at 4 K, see right panel in Fig.~\ref{figSXXX1}c.

%This type-II-like NPL is induced along in-plane direction, see Fig.S\ref{figSXXX1}b,c. However, it behaves likewise the type-II along z direction. Once reached the induced-type-II system, for $Q=1.0$, triplet states become visible, Fig.S\ref{figSXXX1}d.

\begin{figure}
	\centering
		\includegraphics[width=16cm]{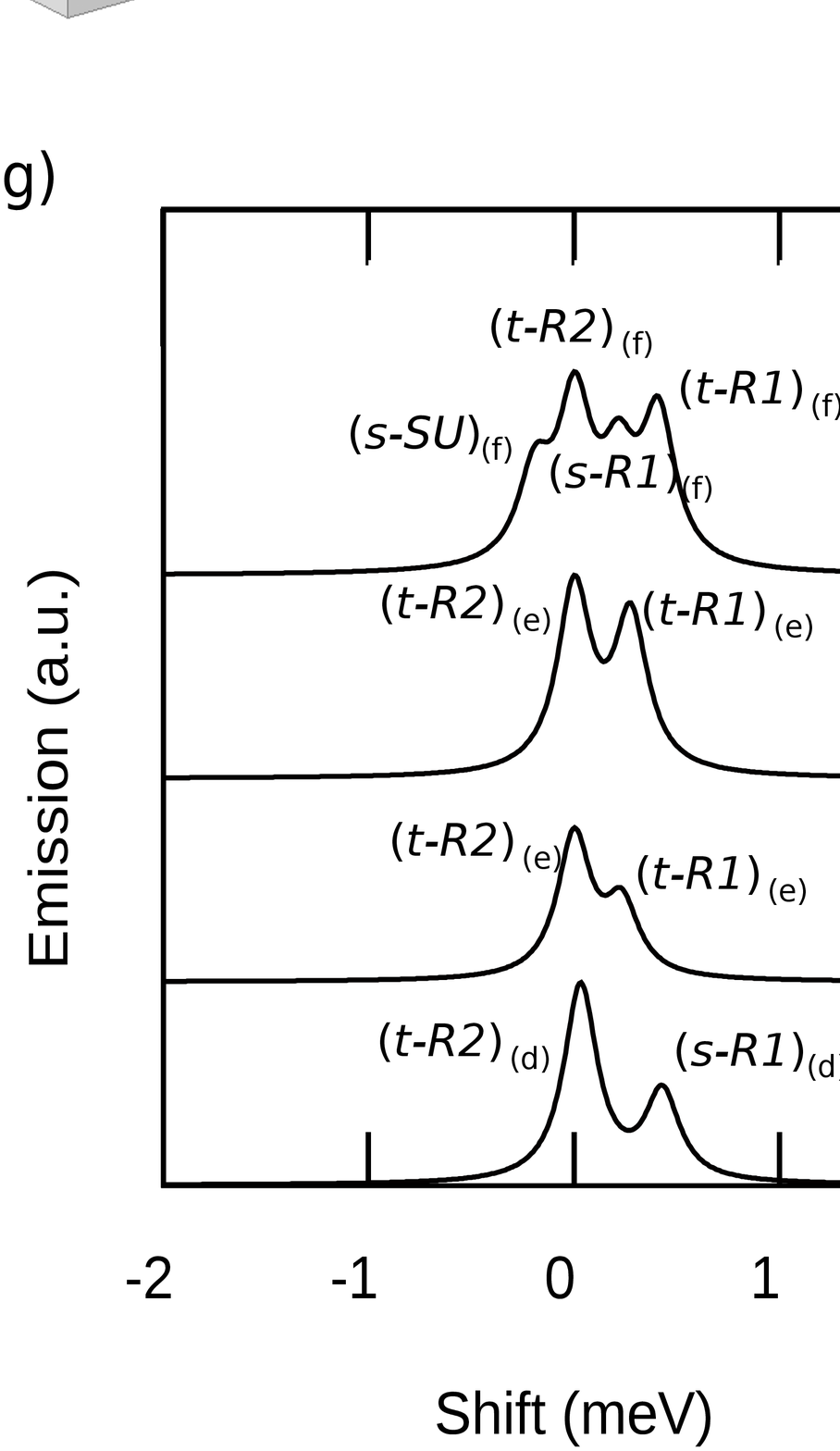}
\caption{(a) Schematic of a CdSe/CdS NPL with 2 double charges on edges intersecting interface and sidewall facet. 
(b,c) Coulomb interactions: (b) electron-electron repulsion and (c) electron-hole attraction for configurations $\ket{1}$ and $\ket{2}$. 
(d-f) Recombination processes involved in each transition. 
(g) Normalized emission spectrum. The energy origin is set at position of the brightest peak, $t$-$R2$. 
$Q_{edge}$ is the net charge on each edge (times the fundamental electron charge).  $ne=nh=12$. }
	%{\bf arregla fig please per a notacio transicions consistent amb main text: s-R1 etc} 
\label{figSXXX2}
\end{figure}

If we further increase the charge $Q$ (e.g. by assuming double point charges on each side of the NPL, see Fig.~\ref{figSXXX2}a),
additional peaks start showing up in the emission spectrum, which is shown in Fig.~\ref{figSXXX2}g.
The sketches in Fig.~\ref{figSXXX2}e-f assign each peak to a corresponding recombination process.
Two transitions come from the X$^-$ singlet ground state, namely its band edge ($s$-$R1$) and first SU ($s$-$SU$) recombinations.
The other transitions are fully radiative recombinations arising from the triplet state, $t$-$R1$ and $t$-$R2$.
The picture is analogous to that proposed in the Discussion section of the main text to explain the multi-peaked emission of
Ref.~\cite{AntolinezNL}, but in this case the triplet state is thermally occupied at 4 K, so there is no need to assume
slow spin relaxation. The top panel in Fig~\ref{figSXXX2}g qualitatively resembles the clusters of four peaks often
observed by Antolinez and co-workers in their photoluminescence measurements\cite{AntolinezNL}, 
although the inter-peak energy splittings here are one order 
of magnitude smaller. As mentioned in the main text, assuming the triplet state is metastable even if it is beyond $kT$
from the singlet ground state, and varying trapped charge location, it may be possible to retrieve the experimental spectra.

